\documentclass[pre,aps,showpacs,supergroupedaddress,epsfig,amsmath,amssymb,eqsecnum]{revtex4}
\usepackage{epsfig,amsmath,amssymb,bm,epsf,graphics,psfrag,verbatim,subfigure,framed}
\usepackage[usenames,dvipsnames]{color}
\usepackage{bbm}
\usepackage{mathrsfs}
\usepackage{amsfonts}
\usepackage{color}
\usepackage{pbox}
\newcommand{\be}{\begin{equation}}
\newcommand{\ee}{\end{equation}}
\newcommand{\ba}{\begin{eqnarray}}
\newcommand{\ea}{\end{eqnarray}}
\newcommand{\bw}{\begin{widetext}}
\newcommand{\ew}{\end{widetext}}

\begin{document}

\title{Modulation of elasticity and interactions in charged lipid multibilayers: \\monovalent salt solutions}
\author{Bing-Sui Lu$^{1,2}$}
\email{binghermes@gmail.com}
\author{Santosh Gupta$^{3,4}$, Michal Beli\v{c}ka$^{3,4}$}
\author{Rudolf Podgornik$^{1,5}$}
\email{rudolf.podgornik@fmf.uni-lj.si}
\author{Georg Pabst$^{3,4}$}
\email{georg.pabst@uni-graz.at}
\affiliation{$^{1}$Department of Theoretical Physics, Jo\v{z}ef Stefan Institute, 1000 Ljubljana, Slovenia.}
\affiliation{$^{2}$School of Physical and Mathematical Sciences, Nanyang Technological University, 21 Nanyang Link, 637371 Singapore.}
\affiliation{$^{3}$University of Graz, Institute of Molecular Biosciences, Biophysics Division, NAWI Graz,Humboldtstr. 50/III, A-8010 Graz, Austria.}
\affiliation{$^{4}$BioTechMed-Graz, A-8010 Graz, Austria.}
\affiliation{$^{5}$Department of Physics, Faculty of Mathematics and Physics, University of Ljubljana, Jadranska ulica 19, SI-1000 Ljubljana, Slovenia.}

\date{\today}


\begin{abstract}
We have studied the electrostatic screening effect of NaCl solutions on the interactions between anionic lipid bilayers in the fluid lamellar phase using a Poisson-Boltzmann based mean-field approach with constant charge and constant potential limiting charge regulation boundary conditions. The full DLVO potential, including the electrostatic, hydration and van der Waals interactions, was coupled to thermal bending fluctuations of the membranes via a variational Gaussian {\em Ansatz}. This allowed us to analyze the coupling between the osmotic pressure and the fluctuation amplitudes and compare them both simultaneously with the measured dependence on the bilayer separation, determined by the small-angle X-ray scattering experiments. High-structural resolution analysis of the scattering data revealed no significant changes of membrane structure as a function of salt concentration. Parsimonious description of our results is consistent with the constant charge limit of the general charge regulation phenomenology, with fully dissociated lipid charge groups, together with a four-fold reduction of the membranes' bending rigidity upon increasing NaCl concentration.
\end{abstract}

\maketitle

\section{Introduction}

Lipid bilayers are well-established mimics of biological membranes, enabling the application of an array of biophysical techniques to study their physicochemical properties.~\cite{Pabst.2014}  Significant efforts have been devoted to unraveling the interactions between adjacent membranes, which are remarkably similar to those between other biological macromolecules or indeed between colloids in general.~\cite{Podgornik.2003} Rigid uncharged membranes are well-described within the Derjaguin-Landau-Verwey-Overbeek (DLVO) paradigm where the total interaction potential can be decomposed into an attractive van der Waals (vdW) part and a repulsive hydration interaction part, respectively ~\cite{Parsegian.1995}, augmented by a short-range steric contribution arising from lipid headgroup collisions of adjacent bilayers at vanishing separations~\cite{McIntosh.1987b}. Both, the vdW as well as the hydration interactions are ubiquitous and not specific for membrane-membrane interactions, as is sometimes claimed for the latter.~\cite{LeiteRubim.2016} Hydration interaction in fact represents a universal, solvent-mediated interaction in a highly structured solvent such as water, observed to occur at small spacings even between completely rigid surfaces and can thus not be ascribed to thermally excited protrusions.~\cite{Stanley.2011}

Membrane charging may occur, e.g., due to (de)protonation or ion-adsorption to lipid headgroups, conferring in principle a long-range electrostatic (ES) interactions to the full membrane-membrane interaction potential~\cite{Cevc.1990}. Membrane electrostatics is typically formulated on the Poisson-Boltzmann (PB) mean-field level~\cite{Markovich.2016}, which entails severe approximations and has in general a well recognized limited range of validity.~\cite{Naji.2013} One of the central results of the PB theory is the existence of salt-ion induced Debye screening, making repulsive electrostatic interactions between symmetrically charged membranes short(er) ranged. However,  the PB predictions can sometimes fail even {\it qualitatively} for physically interesting situations involving highly charged membranes, or multivalent mobile ions, engendering electrostatic interactions between symmetrically charged surfaces that can turn attractive, defying the common wisdom about ES interactions. In what follows we will nevertheless assume the validity of the PB approach in the case of a monovalent salt, NaCl in this case, and even more, rely on the linearization {\sl Ansatz} of the PB equation that allows us to use analytic electrostatic interaction potentials. Furthermore, deprotonation and/or ion-adsorption of solution ions onto the dissociable lipid headgroups, in general leads to the emergence of {\em charge regulation}~\cite{BenYaakov.2011b}, a shorthand for a variable membrane surface charge density that responds to the changes in pH, salt concentration and the separation between membranes.~\cite{Ninham.1971} Charge regulation formally implies different boundary conditions for the membrane electrostatic field \cite{Markovich:2016}, making it dependent on the surface electrostatic potential. While the general description and formal solution of the charge regulated PB theory is complicated~\cite{Behrens.1, Behrens.2, Podgornik.1995}, it possesses two well defined, universal and simple limits that in many cases reduce to either a constant charge (CC) or a constant potential (CP) boundary condition.~\cite{Markovich.2016} 

At finite temperatures fluid membranes exhibit thermally excited bending fluctuations. When constrained by the vicinal bilayers in a membrane stack, inducing steric repulsions between colliding membranes, these fluctuations lead to entropic long-range repulsive interactions, first proposed and formalized by Helfrich.~\cite{Helfrich.1978} In fact, bending fluctuations not only turn short-range contact steric repulsions into long-range Helfrich interaction, but also thermally renormalize other soft DLVO interactions, such as the ES and the vdW interactions, which in this context we refer to as the {\sl bare interactions} \cite{Evans.1986, Podgornik.Parsegian.1992}. In some limiting cases the effect of this thermal renormalization can be approximately captured by adding an additional long-ranged entropic thermal potential of a Helfrich type to the underlying DLVO bare interactions~\cite{Lu.2015}. However, this approach has severe limitations and generally one needs to either develop a more sophisticated theoretical approach that takes fully into account the coupling between the bare interactions and their thermal renormalization~\cite{Lipowski:book}, or to investigate the thermal effects by performing extensive numerical simulations starting with an assumed form of the underlying bare interaction potentials \cite{Gouliaev.1998}. The latter approach was successfully implemented in Fourier-mode Monte Carlo simulations of a stack of fluctuating charge neutral membranes~\cite{Gouliaev.1998b} and was recently, among other things, successfully applied to the problem of fluid-fluid phase separation in membranes.~\cite{Kollmitzer.2015b} 

While the simulation approach is in many respects the preferred one, it entails usually inordinate computational times if one wants to fit experimental data and deduce from the fits effective interaction parameters such as the Hamaker coefficient in the case of vdW interactions, or the effective surface charge density/surface potential in the case of ES interactions. As a viable alternative for these purposes we recently developed a simplified analytical theory of thermally renormalized effective interactions between fluid membranes, based on a generalized Gaussian variational {\sl Ansatz}, that takes into account the coupling between membrane fluctuations and the underlying bare interaction potentials through the application of the Feynman-Kleinert variational field theory~\cite{Lu.2015}. In this respect it represents an update on previous attempts based on the Gaussian variational {\sl Ansatz}~\cite{Podgornik.Parsegian.1992, Odijk, deVries1,deVries2,deVries3}. This approach was shown to provide numerically tractable analytical results, being thus ideally suited for fitting experiments that provide concurrently the osmotic pressure, i.e. the osmotic equation of state for a stack of membranes, {\em as well as} the strength of membrane shape fluctuations as experimental data. The osmotic equation of state simply connects the membrane volume fraction or the separation between membranes in a stack, with the applied osmotic pressure set by a standardized concentration variation of an osmolyte,  such as polyethylene glycol (PEG) \cite{Cohen.2009}. We will therefore exploit the approximate analytical theory of thermally renormalized effective inter-membrane interactions in order to deduce the bare interaction parameters from a {\em coupled fit} to both the measured osmotic equation of state as well as the membrane bending fluctuations. This coupled fit is much more restrictive than the usual fitting procedure for interaction parameters, that relies on the equation of state only, and actually creates much more stringent demands on the realism and consistency of the theoretical description then the single data fit.

Previously, the interplay between thermal undulations and electrostatic repulsion has been analyzed in terms of the shifts of the Bragg peak and back-scattering in light-scattering experiments on dilute lamellar phases of the nonionic surfactant n-dodecyl pentaethylene glycol ether, with small added amounts of the ionic surfactant sodium dodecyl sulphate \cite{deVries2}, or dilute lamellar phases of the non-ionic surfactant triethylene glycol monodecyl ether, with small amounts of the anionic surfactant sodium dodecyl sulphate (SDS), both with and without added electrolyte \cite{deVries3}. Furthermore, small angle neutron scattering and neutron reflectivity of lamellar phases containing pentaethylene glycol n-dodecyl ether, sodium decylsulfonate, and $\rm D_2O$ was measured at the solid-liquid interface as the molar ratio of ionic to nonionic surfactant is changed and compared with theoretical predictions \cite{deVries4}. However, these experiments were not set up to probe the swelling behavior of surfactants with controlled osmotic pressure concurrently with the fluctuations of the multilamellar systems. Such experiments were performed for charged lipid and surfactant systems, however, without considering the renormalization of bare interactions due to thermal fluctuations of the interacting  membranes.~\cite{Cowley:1978, LoosleyMillman.1982, McIntosh.1990,Brotons.2005b, Brotons.2006}

For charge neutral lipid bilayers Petrache and coworkers in a series of papers focused on the modulation of membrane interactions by ion-specific effects.~\cite{Petrache.2005, Petrache.2006b, Petrache.2006} Performing osmotic stress experiments, similar to those applied in the present study, they observed a screening effect of the monovalent salt on the vdW interactions at high salt, specifically on the zero frequency term of the Matsubara sum \cite{Parsegian.2004}, while no changes were observed in the membrane structure or indeed its bending rigidity. This should expectedly not be the case for charged membranes, considered in the present study, where the affinity of ions for charged membranes is significant, thus potentially affecting not only the interactions between the membranes but also single-membrane elasticity as well as membrane structure. In particular, theoretical expectations would imply a larger bending rigidity for charged membranes~\cite{Andelman.2006}, which might be dependent on the Debye screening and the type of solution ions. 

In what follows we will focus on dipalmitoyl phosphatidylglycerol (DPPG) bilayers in NaCl solutions. Phosphatidylglycerols are abundant in mitochondria~\cite{vanMeer.2008} or bacterial plasma membranes~\cite{Lohner.2008} and are fully deprotonated at neutral pH, with a pK$_a$ of 2.9.~\cite{Watts.1978}. The structural properties of DPPG bilayers, as well their thermotropic behavior are well-documented.~\cite{Zhang.1997, Pabst.2007b, Pan.2014b} By developing an analytic theory of thermally renormalized DLVO interactions, coupling all bare interactions to membrane bending fluctuations, we were able to derive and parameterize the effective interactions between anionic DPPG and their modulation by salt from osmotic stress experiments. ES interactions were modeled in terms of the charge regulation theory in the limits of fully dissociated ions and an equilibrium of bound and dissociated states. Comparison with experiments showed that it is the former limiting case that is realized in the physical system under investigation. Moreover our results emphasize a previously unrecognized large contribution of bending fluctuation driven steric repulsion between bilayers, which becomes increasingly screened by the salt content, while we find no significant change of the surface charge that goes along with ion concentration. This is consistent with an unexpectedly weak Na$^+$ binding to the charged headgroups.

\section{Materials and Methods}
\subsection{Experimental}
\subsubsection{Sample Preparation}
DPPG was purchased from Avanti Polar Lipids (Alabaster, AL) and used without further purification. NaCl was obtained from Karl Roth (Karlsruhe, Germany) and polyethylene glycol (PEG, $M_W$ = 8000) was purchased from Sigma-Aldrich (Vienna, Austria). Lipid stock solutions were prepared by dissolving predetermined amounts of dry lipid in chloroform/methanol (9:1, v/v) and subsequently dried first under a stream of nitrogen and then under vacuum for about 12h to form a thin lipid film on the bottom of glass vials. To achieve positionally correlated multibilayers (i.e. free-floating multilamellar vesicles) with equal concentrations of ions in the interstitial water layers, we first prepared unilamellar vesicles by hydrating the lipid films in HEPES buffer (10 mM, pH 7.4), followed by one hour vortex-mixing by using a MaxQ benchtop shaker (Thermo Fischer Scientific,  Waltham, MA) at 500 rpm and 30 min ultra-sonication in a water bath. Then appropriate amounts of concentrated NaCl solutions in HEPES buffer were added (resulting in bulk NaCl concentrations ranging from 100 mM -- 700 mM NaCl) and the sample was again vortex mixed for one hour. 

In a typical osmotic pressure experiment, the samples were cooled to room temperature and transferred to conically shaped test tubes for 1 hour centrifugation at 15000 rpm. After removing the supernatant, the pellets were overlaid with PEG dissolved at appropriate concentration in HEPES/NaCl buffer. PEG concentration varied from  1--42 wt\%, corresponding to osmotic pressures of 0.025--35 atm.~\cite{Stanley.2003} After overlaying samples with argon for protection against oxidation, the vials were closed, taped, and stored at room temperature for 7$-$10 days prior to measurement.

\subsubsection{Osmotic Stress Experiments}
Small-angle X-ray scattering (SAXS) experiments were  performed at the Austrian SAXS beamline at ELETTRA, Trieste, Italy, using 8 keV photons at  an energy dispersion, ${\Delta E}/E$, of 2.5x10$^{-3}$. A mar300 image plate detector (marresearch,Norderstedt, Germany) covering a $q$-range from 0.2 to 0.7 \AA$^{-1}$ was used and calibrated for scattering angles using silver-behenate. For data acquisition, the samples were filled into reusable quartz-glass capillaries (diameter: 1 mm) and kept in a brass sample holder connected to a circulating water bath (Huber, Offenburg, Germany). Samples were equilibrated for 10 min before exposing them for 30 s to the X-ray beam.

The two-dimensional detector signal was radially integrated using FIT2D and corrected for scattering contributions from the buffer and from the capillary. For osmotically stressed samples additional scattering from PEG made a standard background subtraction impractical. Since the essential information in this case are the shape and the positions of Bragg peaks, we subtracted approximate backgrounds, obtained by interpolating between SAXS signals of NaCl in HEPES buffer and PEG/NaCl in HEPES buffer mixtures. Alternatively, one could just subtract an arbitrary smooth function from the measured patterns. 


\subsubsection{Analysis of Scattering Data}

We applied the full global data analysis previously detailed by Heftberger et al.\cite{Heftberger.2014} In brief, the scattering intensity of unoriented multibilayers is of the form
\begin{equation}
I(q) = \frac{1}{q^{2}} [|F(q)|^{2} S(q) (1-N_{diff}) + |F(q)^{2}| N_{diff}]
\end{equation}
where $q$ is the scattering vector,  $S(q)$ is the structure factor,  $F(q)$ is the form factor and $N_{diff}$ accounts for diffuse scattering originating from positionally uncorrelated bilayers. The structure factor accounts for positional correlations within the multibilayers and is a function of the lamellar repeat distance $d$, the average number of layers per scattering domain and the Caill\'{e}/bending fluctuation parameter $\eta$, describing the line-shape of the Bragg peaks.~\cite{Zhang.1994,Pabst.2000}  From the $\eta$ parameter, the mean square fluctuations of the bilayer separations can be calculated using $\Delta^2_{{\rm exp}}=\eta d^2/ \pi^2$.~\cite{Petrache.1998b}.

The form factor is the Fourier transform of the single bilayer electron density profile, which we modeled in terms of the scattering density profile (SDP) model.~\cite{Kucerka.2008} The model parses the bilayer lipids into quasi-molecular fragments in terms of volume probability distributions. In particular,  we followed the reported parsing for phosphatidylglycerols, describing the bilayer structure by the (i) terminal methyl (CH3), (ii) methylene (CH2), (iii) carbonyl glycerol (CG) (iv) phosphate (PG1) and (v) glycerol (PG2) groups (see also Fig.~\ref{fig:fit_parse_ed}B in the results section).~\cite{Pan.2014b}


Membrane structural parameters, such as, e.g. area per lipid molecule $A$, were defined and calculated from the SDP profiles as described previously.~\cite{Heftberger.2014}. Following the work by Kollmitzer and coworkers on domain interactions,~\cite{Kollmitzer.2015b} we focus in particular on the steric membrane thickness,~\cite{McIntosh.1993} which is defined as the distance between the remotest atoms of the lipid molecule, $d_{B}^{S}=2(z_{PG2}+\sigma_{PG2})$, where $z_{PG2}$ and $\sigma_{PG2}$ are the position (measured from the bilayer center) and the width of the glycerol group in the lipid headgroup, respectively. The average thickness of water layer (bilayer separation) is given by $\langle \ell_0 \rangle = d-d_{B}^{S}$.       

\subsection{Theory}

The model used to analyze the osmotic stress data is physically represented by a pair of interacting fluid membranes whose mean positions are co-planar. Each membrane can undergo thermally driven undulations about its mean position. Further, the mean membrane position is also free to undergo thermal fluctuations. The problem of a two-membrane system can be transformed to that of a membrane fluctuating with a hard wall (cf. Supporting Information), with an effective bending energy
\begin{equation}
H_{b} = \frac{K_{{\rm eff}}}{2} \int \! d^2\mathbf{x}_\perp \, (\nabla^2 \ell(\mathbf{x}_\perp))^2,
\end{equation}
where $\mathbf{x}_\perp=(x,y)$ is a two-dimensional coordinate on the transverse projected plane of the membranes, $\ell(\mathbf{x}_\perp)$ is the interbilayer separation between the two membranes, and $K_{{\rm eff}} \equiv (K_1 K_2)/(K_1 + K_2)$ is the ``effective''  bending rigidity. For similar membranes, $K_1=K_2\equiv K$, and $K_{{\rm eff}}=K/2$. 
In the above, we have neglected nonlinear gradient coupling terms  that are less important than the coupling terms stemming from the interaction potential.~\cite{Lipowsky.1986}

Furthermore we decompose the intermembrane separation into two contributions
\begin{equation}
\ell(\mathbf{x}_\perp) = \ell_0 + \delta \ell(\mathbf{x}_\perp), \qquad {\rm with} \qquad \int \! d^2\mathbf{x}_\perp \ell(\mathbf{x}_\perp) = S ~\ell_0
\end{equation}
where $\ell_0$ is the (instantaneous) \emph{geometric} mean separation (or \emph{rigid} bilayer separation) between the two membranes and $S$ is the transverse projected area of each membrane. $S$ is coupled to membrane undulations, i.e. large fluctuations lead to a strongly crumpled membrane surface and a reduced the value of $S$.

\subsubsection{Interaction Potentials}

Membrane fluctuations are not just influenced by steric interactions between membranes, but also by soft long(er)-ranged non-steric interactions such as hydration, ES and vdW interactions, so that both steric and non-steric interactions contribute to the fluctuation-induced osmotic stress that the membranes experience \cite{Safran.2003}.  Hence, in addition to bending fluctuations, the effective mesoscopic energy of the considered system also contains contributions from bare interactions and externally applied osmotic pressure $P$
\begin{equation}
H = H_b + H_I + S \ell_0 P, 
\label{bfcsjyer}
\end{equation}
where 
\begin{equation}
H_I = \int \! d^2\mathbf{x}_\perp \, (V_{h} + V_{es} + V_{vdw}) = \int \! d^2\mathbf{x}_\perp \, w(\ell(\mathbf{x}_\perp)).
\label{eq:HI}
\end{equation}
originates from the interaction potential and contains  contributions from the hydration potential, $V_h$, the ES interaction energy, $V_{es}$, and the vdW interaction energy, $V_{vdw}$ (all normalized to unit area). These are given standardly by \cite{Safran.2003}
\begin{equation}
V_h(\ell) = P_H \lambda_H e^{-\frac{\ell}{\lambda_H}}; 
\end{equation}
\begin{equation}
V_{vdw}(\ell) = - \frac{W \, g(\ell_0)}{12\pi}\Big[ \frac{1}{\ell^2} - \frac{2}{(\ell+d_{{\rm B}})^2} + \frac{1}{(\ell+2d_{{\rm B}})^2} \Big], 
\end{equation}
where $P_H$ is the hydration pressure, $\lambda_H$ is the hydration decay length, $W$ is the Hamaker coefficient, $d_B$ is the bilayer thickness (which we define via the steric thickness), and $g(\ell_0)$ is a cut-off function that reflects the fact that the vdW interaction cannot be singular at zero inter-bilayer separation (see Supporting Information, Eq. (3)).~\cite{Podgornik.2004} The electrostatic interactions depend crucially on the nature of the membrane charges \cite{Markovich.2016}, that result from dissociable molecular moieties through the Ninham-Parsegian  charge regulation process.~\cite{Ninham.1971} Often one assumes that the charge regulation can be approximated by either CC or  CP boundary conditions (BC), though this is not valid in general (for details see~\cite{Markovich.2015}). Constant charge BC reflect complete ion dissociation from the membrane, whilst constant potential BC reflect a state where ions can dissociate and re-associate between the solution and the membrane lipid headgroups. The electrostatic interaction per unit area for the two extreme cases of BCs is approximately~\cite{Parsegian.1972, BenYaakov.2010, Markovich.2016}: 
\begin{eqnarray}
V_{es,{{\rm CC}}}(\ell) &=& P_{es} \lambda_D (\coth(\ell/2\lambda_D)-1)  \qquad\quad (\text{constant charge BC})
\label{eq:VeCC}
\\
V_{es,{{\rm CP}}}(\ell) &=& -P_{es} \lambda_D ( \tanh(\ell/2\lambda_D) - 1 ) \qquad\quad (\text{constant potential BC})
\end{eqnarray}
where the Debye screening length $\lambda_D = (4\pi \ell_B I)^{-1/2}$, with $\ell_B \sim 0.74$ nm being the Bjerrum length and $I \equiv 2 c_{b}$ is the ionic strength, with  $c_b$ being the bulk concentration of salt. $P_{es} \equiv \sigma_s^2/\epsilon_W\epsilon_0$, where $\epsilon_0 \approx 8.85\times 10^{-12}$ F/m  is the vacuum permittivity, $\epsilon_W\approx 69.9$ is the relative permittivity of the water medium at  $T=50^\circ \rm{C}$, and $\sigma_s$ is the surface charge density of the membrane for the CC boundary condition and $\sigma_s = -\epsilon_0\epsilon_W\psi_s/\lambda_D$ for the CP boundary condition, with $\psi_s$ being the surface potential. 
 
Previously we have shown that the thermodynamic properties of the system defined by Eq. \ref{bfcsjyer} can be handled by first deriving  a ``steric potential",  followed by applying the Feynman-Kleinert variational field theory to account for contributions of the steric and non-steric interactions.~\cite{Lu.2015} Specifically, the steric potential is derived by (i) implementing the steric constraint at the level of the partition function and (ii) promoting this constraint into an energetic term via an analytic representation of the Heaviside function in the saddle-point approximation. The steric potential in the absence of non-steric interactions leads directly to the same separation dependence evinced by the Helfrich interaction,~\cite{Helfrich.1978, Helfrich.1984} but with an additional dependence on temperature and bending rigidity as compared to previous treatments.~\cite{Podgornik.1992}  The variational approach reflects the idea that the overall free energy of the interacting membrane system \emph{cannot} be given by an additive sum of a ``bare" steric potential and non-steric interaction terms. In the variational framework, the non-additivity is captured by first solving a variational interaction-dependent equation for the mean-square membrane undulation, then feeding the solution into the formula for the osmotic pressure predicted by theory. 

The basic quantity in the self-consistent variational framework is the variational free energy per unit area, which we derive to be 
\begin{eqnarray}
f_{var}(\ell_0, \Delta_u) =  w_u(\ell_0) + \frac{(k_{{\rm B}}T)^2}{128 \, K_{{\rm eff}} \, \Delta_u^2} + \frac{9 k_{{\rm B}}T c^2 \Delta_u^2}{8\ell_0^4} + \frac{3c^2 k_{{\rm B}}T}{8 \ell_0^2} 
+ P_{\rm{osm}} \ell_0,
\label{narlewi}
\end{eqnarray}
where $c = \lambda_s / a$ is the so-called penetration coefficient reflecting the fact that opposing membranes could penetrate into each other over a narrow length scale when brought in close contact due to the softness of the bilayer surface, with $\lambda_s$ being the length of the steric penetration layer and $a$ being the diameter of the lipid headgroup in the plane of the lipid bilayer.~\cite{Lu.2015} Physically, the region that the fluctuating membrane can access (i.e., the aqueous interbilayer region) is separated from the region that it cannot access (i.e., inside the bilayer, which is the ``wall") by a plane. This plane is not infinitely thin, but rather has a certain molecular-order thickness $\lambda_s$ which is of the order of the size of the lipid headgroup. Depending on the molecular make-up the steric thickness can differ from the size of the headgroup by some factor, and $c$ quantifies this difference. $\Delta_u^2$ is the variational Gaussian approximation to the mean square undulation $\langle \delta  \ell^2 \rangle$ of the membrane~\cite{Lu.2015}, and $w_u$ corresponds to a variational approximation for the bare interaction terms, i.e., hydration, ES and vdW potentials, renormalized by thermal fluctuations
\begin{equation}
w_u(\ell_0) \equiv \int_{-\infty}^{\infty}\!\!\!\frac{d\ell}{\sqrt{2\pi\Delta_u^2}} w(\ell)\,e^{-\frac{(\ell-\ell_0)^2}{2\Delta_u^2}} 
\label{eq:w_var}
\end{equation}
(see Supporting Information for details).  In order to arrive at the full fluctuation spectrum, as measured by experiment, we need to include ``zero mode'' fluctuations, reflecting the positional variations of rigid bilayers. This is achieved by making use of the decomposition $\ell = \langle \ell_0 \rangle + \delta\ell_0 + \delta\ell$, where $\langle \ell_0 \rangle$ is the equilibrium rigid bilayer separation and $\delta\ell_0$  the thermal fluctuation of the rigid bilayer separation. Note that the averaging $\langle \ldots \rangle$ is taken with respect to a thermal equilibrium ensemble of undulations and fluctuating rigid bilayer separations. Then the square of all thermal fluctuations  can be written as
\begin{equation}
\Delta_{{\rm full}}^2 = \langle \delta\ell_0^2 \rangle + \langle \delta  \ell^2 \rangle \simeq \Delta_0^2 + \Delta_u^2, 
\end{equation}
which corresponds to $\Delta_{{\rm exp}}^2$ defined above. The variational Gaussian approximation for the zero mode fluctuation amplitude can be found from
\begin{equation}
\label{eq:meansquarefluct}
\langle \delta\ell_0^2 \rangle \simeq \Delta_0^2 = 
\frac{k_{\rm{B}} T}{S} 
\left[ \frac{\partial^2 f_{{\rm var}}(\ell_0, \Delta_u)}{\partial \ell_0^2} \right]^{-1}.
\end{equation}
The higher mode fluctuations in turn are found by minimizing $f_{var}(\ell_0, \Delta_u)$ with respect to $\Delta_u$. This leads to the variational equation
\begin{equation}
\frac{\partial f_{var}(\ell_0, \Delta_u)}{\partial \Delta_u} \equiv 0, 
\end{equation}
wherefrom $\Delta_u = \Delta_u(\ell_0)$. The physical requirement that the mean square membrane fluctuations must not become negative ($\langle \delta\ell_0^2 \rangle + \langle \delta\ell^2 \rangle \geq 0$) provides us with a stringent control of our Gaussian approximation for fluctuations. Since major contributions to $\Delta_{{\rm full}}^2$ originate from the zero mode fluctuations, this is equivalent to requiring that the Hessian of the variational free energy (Eq.~(\ref{narlewi}))
\begin{equation}
\mathcal{H}\equiv \frac{\partial^2 f_{{\rm var}}(\ell_0, \Delta_u(\ell_0))}{\partial \ell_0^2}
\label{gfdyei}
\end{equation}
be non-negative. If this is not the case the Feynman-Kleinert variational scheme is not applicable.

Effectively, this furnishes us with a criterion to determine the range of separations $\ell_0$ and interaction strengths over which the variational Gaussian approximation holds. Specifically we have to reject boundary conditions which yield $\mathcal{H} < 0$ at any membrane separation as this would imply negative mean square fluctuations.
This enables us to discriminate between constant charge and constant potential BC, yielding an insight into the character of the charge dissociation from the membranes.

To model experimental data the osmotic pressure has to be determined for each $\ell_0$ using $\Delta_u(\ell_0)$ by evaluating 
\begin{eqnarray}
\frac{\partial f_{var}(\ell_0, \Delta_u(\ell_0))}{\partial \ell_0} \equiv 0, 
\label{cdafgs}
\end{eqnarray}
taking into account Eq. \ref{narlewi}. For all further excruciating details we direct the reader to reference.~\cite{Lu.2015}

\subsubsection{Analysis of Fluctuations and Osmotic Stress Data}

The sequence of equations of Eqs. \ref{bfcsjyer} -- \ref{cdafgs} presents also the flowchart of the applied fitting procedure and data analysis. Fits were obtained by optimizing $a_\alpha$ (where $\{ a_\alpha\}_{\alpha=1}^{7} \equiv \{ P_H, \lambda_H, W, c, K_{{\rm eff}}, P_{es}, S\}$; cf. Eqs.~(\ref{eq:HI}) and (\ref{narlewi}));  $\lambda_D$ is given by experimental conditions and $d_B$ from the SAXS data analysis as detailed above. 

The optimization is performed by minimizing the function
\begin{equation}
\label{eq:chi-square}
\chi^2(\{ a_\alpha \}) \equiv \gamma \chi_{{\rm P}}^2(\{ a_\alpha \}) + \chi_{{\rm f}}^2(\{ a_\alpha \}), 
\end{equation}
where $\chi_{{\rm P}}^2$ and $\chi_{{\rm f}}^2$ are the minimization functions for $P_{osm}(\ell_0)$ and $\Delta_{full} (\ell_0)$, respectively, and $\gamma$ is a relative weight, which accentuates the global minimum of $\chi_{{\rm P}}^2$. For our purposes $\gamma = 1000$ was sufficient to produce reasonable fits to the osmotic pressure data. Further, we increased for each salt concentration the relative weight of the lowest $P_{osm}$ data by a factor of $1000$ to ensure that the fitted pressure curves follow the trend of the experimental data towards a finite bilayer separation. Due to the large number of fitting parameters simulated annealing was applied as global search algorithm using Mathematica's NMinimize routine.

\section{Results and Discussion}

\begin{figure*}[t!]
		\includegraphics[width=1\textwidth]{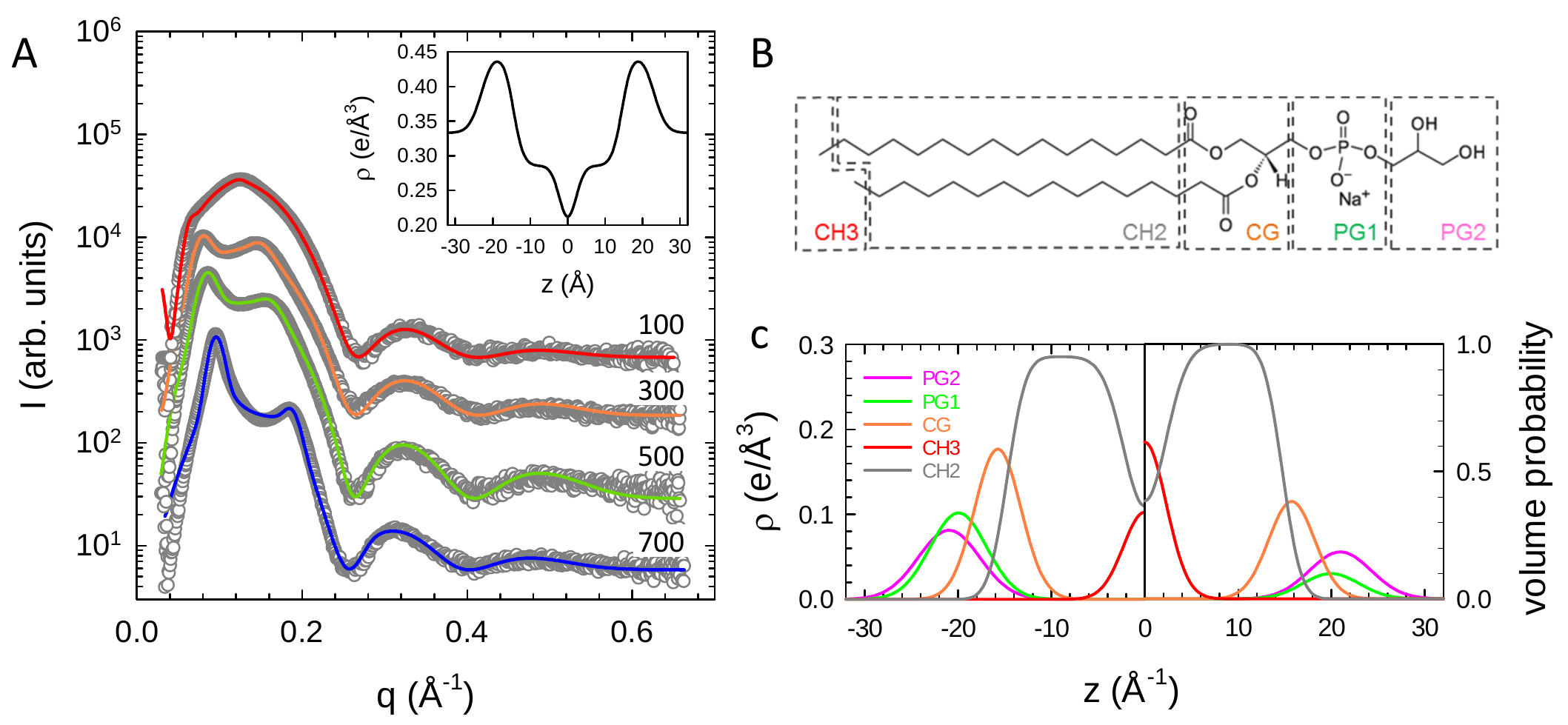}
	\caption{Global analysis of SAXS data in terms of the SDP model. Panel (A) gives the scattering patterns of unstressed DPPG multibilayers at 50$^\circ$C as a function of NaCl concentration. Numbers right to the patterns give salt concentration in mM. For clarity of display data have been offset by a constant. Solid lines correspond to the best fits from the SDP-GAP model. The resulting electron density profile at 700 mM NaCl is shown in the inset. Panel (B) illustrated the applied parsing scheme for DPPG and panel (C) shows the corresponding volume probability distributions (\textit{right hand side}) and electron densities (\textit{left hand side}).}
\label{fig:fit_parse_ed}
\end{figure*}

SAXS patterns of DPPG obtained in the absence of osmotic pressure exhibited Bragg peaks, whose occurrence became more and more prominent with increasing salt concentration (Fig.~\ref{fig:fit_parse_ed}A). In particular peaks became sharper and shifted to higher $q$-values, indicating screening of electrostatic interactions by monovalent salt, which increased positional correlations between the bilayers. All peaks were indexed on a single lamellar lattice, allowing us to apply the full $q$-range analysis detailed in the Materials and Methods section. 
From the analysis we found a decrease of $d$ from 99.5 \AA\/ to 65.7 \AA\/ and of $\Delta_{\rm{exp}}$ from 17.8 \AA\/ to 8.8 \AA\/, respectively, when the NaCl concentration was raised from 100 to 700 mM. These observations can be understood qualitatively in terms of ion-mediated ES screening. Membrane structural parameters did, however, not change significantly with salt concentration. Averaged over all samples we found the area per lipid to be $A = 62.6 \pm 1.3$ \AA$^2$ and $d_B = 38.2 \pm 0.8$ \AA\/ for the Luzzati thickness, defined via the Gibbs dividing surface between the polar and apolar membrane regions.~\cite{Kucerka.2008}  Comparison to literature values obtained by a joint analysis of neutron and x-ray data on DPPG unilamellar vesicles (i.e. in the absence of salt), $A = 64.7 \pm 1.3$ \AA$^2$ and $d_B = 36.7 \pm 0.7$ \AA\/,~\cite{Pan.2014} yields within experimental error reasonable agreement. 
As detailed in the previous section, delineating bilayer interactions requires the steric bilayer thickness, for which we find $d_{B}^{S} =  47.8 \pm 0.6$ \AA\/ as average value over all salt concentrations.

\begin{figure*}[t!]
\includegraphics[width=0.7\textwidth]{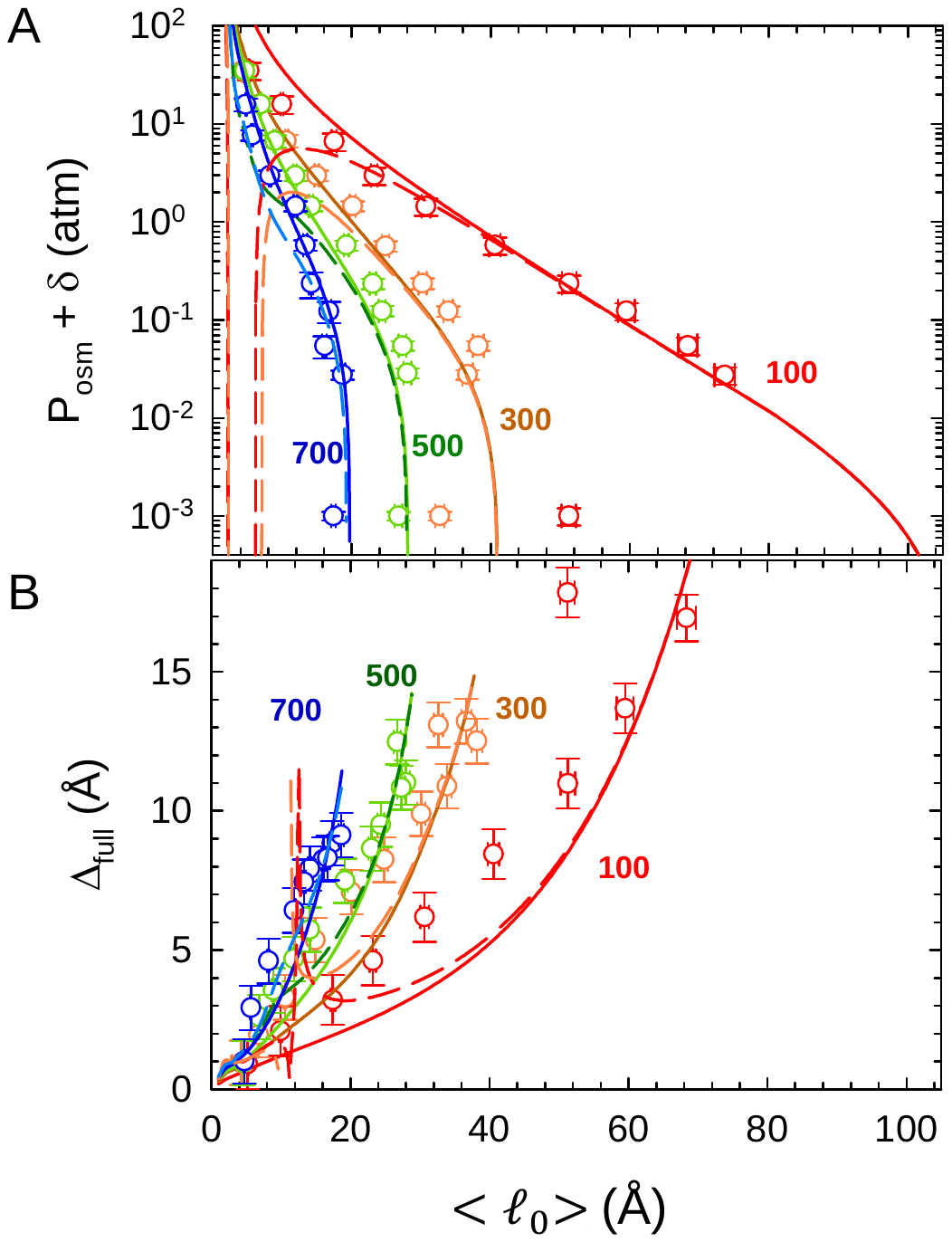}
	\caption{Screening of electrostatic interactions between fluid DPPG bilayers in the presence of NaCl. Panel A gives the osmotic pressure isotherms and panel B the corresponding membrane fluctuation amplitudes as a function of the equilibrium spacing $\langle \ell_0 \rangle$. A constant of $\delta = 10^{-3}$ atm was added to the $P_{osm} $ data to include zero osmotic pressure data on the log-scale. Numbers adjacent to data denote the given salt concentration in mM. Solid lines represent the best fits obtained from CC boundary conditions. The dashed lines correspond to the solutions obtained using CP boundary conditions. Resulting fit parameters for CC boundary conditions are displayed in Tab.~(\ref{table:tablefluct1}) and Fig.~(\ref{fig:params}).}

\label{fig:flucts}
\end{figure*}

In order to gain quantitative insight on the ion-mediated modulation of bilayer interactions we performed osmotic stress experiments. In general, application of osmotic pressure lead to a decrease of bilayer separation and fluctuations (Fig.~\ref{fig:flucts}) as observed in several osmotic stress experiments previously (see, e.g.~\cite{Parsegian.1995, Petrache.1998b, McIntosh.2000, Pabst.2007, Kollmitzer.2015b}), including studies on highly charged surfactants~\cite{Brotons.2005b,Brotons.2006}.

Interactions in the currently studied bilayer separation ranges can be roughly divided into two regimes. At low hydration/high $P_{osm}$ or equivalently small $\langle \ell_0 \rangle$, repulsive hydration forces dominate, while interactions at lower $P_{osm}$ or equivalently larger $\langle \ell_0 \rangle$ are characterized by a balance of repulsive ES and steric fluctuation interactions and attractive vdW interactions. 
Eventually, at low enough $P_{osm}$ or equivalently largest $\langle \ell_0 \rangle$, vdW interactions become dominant, leading to a finite $\langle\ell_0\rangle$-value at which $P_{osm} = 0$, i.e., the repulsive and attractive interactions are compensated and the system swells into a well defined secondary DLVO-like minimum. Swelling of the DPPG multibilayers became more and more pronounced with decreasing NaCl concentration, typically because of a reduced screening of ES interactions (Fig.~\ref{fig:flucts}A). At the same time bilayer fluctuation amplitudes increased (Fig.~\ref{fig:flucts}B), reflecting the increased separation between the bilayers, modified bilayer properties, and/or changed bilayer interactions. 

Intriguingly, all samples - including several replicates - exhibited smaller bilayer separations in the completely unstressed state, defined by $P_{osm} = 0$, than at the smallest but finite applied osmotic stresses. This completely reproducible effect became increasingly prominent with decreasing salt concentration and amounted to a difference in spacing of up to about 20 \AA\/ for 100 mM NaCl. To the best of our knowledge this anomaly has not been clearly reported and discussed previously and its origin seems to be difficult to pinpoint directly. At this point we can only speculate that unilamellar vesicles, which will peel off during sample preparation and actual measurement, could constitute another osmotically active component of the solution, exerting its own {\sl vestigial osmotic pressure} in addition to the one controlled by the osmotically active dissolved PEG. The final spacings of the multilamellar subphase would then be governed by the sum of the direct PEG osmotic pressure, under full experimental control, and the vestigial osmotic pressure, contingent on the method of the sample preparation. By its very nature the action of the vestigial osmotic pressure is therefore to increase the actual osmotic pressure and/or decrease the interlamellear spacings whose effects could only be observed when the direct osmotic stress of PEG is small and/or vanishing. The lipid-peeling off effect that would constitute the basis for the vestigial osmotic stress is possibly also a non-equilibrium phenomenon, complicating additionally its straightforward quantification.

In other words, the nominal and the actual osmotic pressures in our samples are not the same, the latter incorporating also the vestigial osmotic pressure of the peeled-off vesicles floating freely in solution. It thus seems reasonable to assume that the weakly bound systems, such as those corresponding to low salt concentration, would have increased amounts of such vesicles due to their increased bending fluctuations that in their turn promote more vigorous peeling off. While the fluctuation-mediated membrane unbinding has been discussed previously ~\cite{Pabst.2002, PozoNavas.2003}, however, deriving a reliable measure of the magnitude of this vestigial, possibly non-equilibrium, osmotic pressure in combination with an analytical treatment is out of the scope of the present report and might be in general quite difficult to come by.

As a consequence, in our osmotic stress analysis detailed in the next paragraphs, we have omitted the $P_{osm}=0$ data for 300 and 100 mM NaCl concentrations, where these anomalous effects are most pronounced. Omitting the $P_{osm} = 0$ data had the immediate consequence that the unconstrained global optimization approach detailed in the previous section failed to produce physically realistic Hamaker coefficients. Henceforth, we decided to apply an eye-guided fit of the osmotic stress data - as opposed to the $\chi^2$-minimization procedure - not that different from what was usually adopted previously by other authors.~\cite{Petrache.1998b, Petrache.2006b} On the other hand, where it is possible to perform a $\chi^2$-minimization procedure, as in the fluctuation data, we have used the procedure to estimate the parameter that affects only on the fluctuation amplitude, viz., the membrane area $S$ (cf. Supporting Information, Eq.~(17)). Reported errors are estimates from the unconstrained fits.

\begin{figure*}[t!]
		\includegraphics[width=0.6\textwidth]{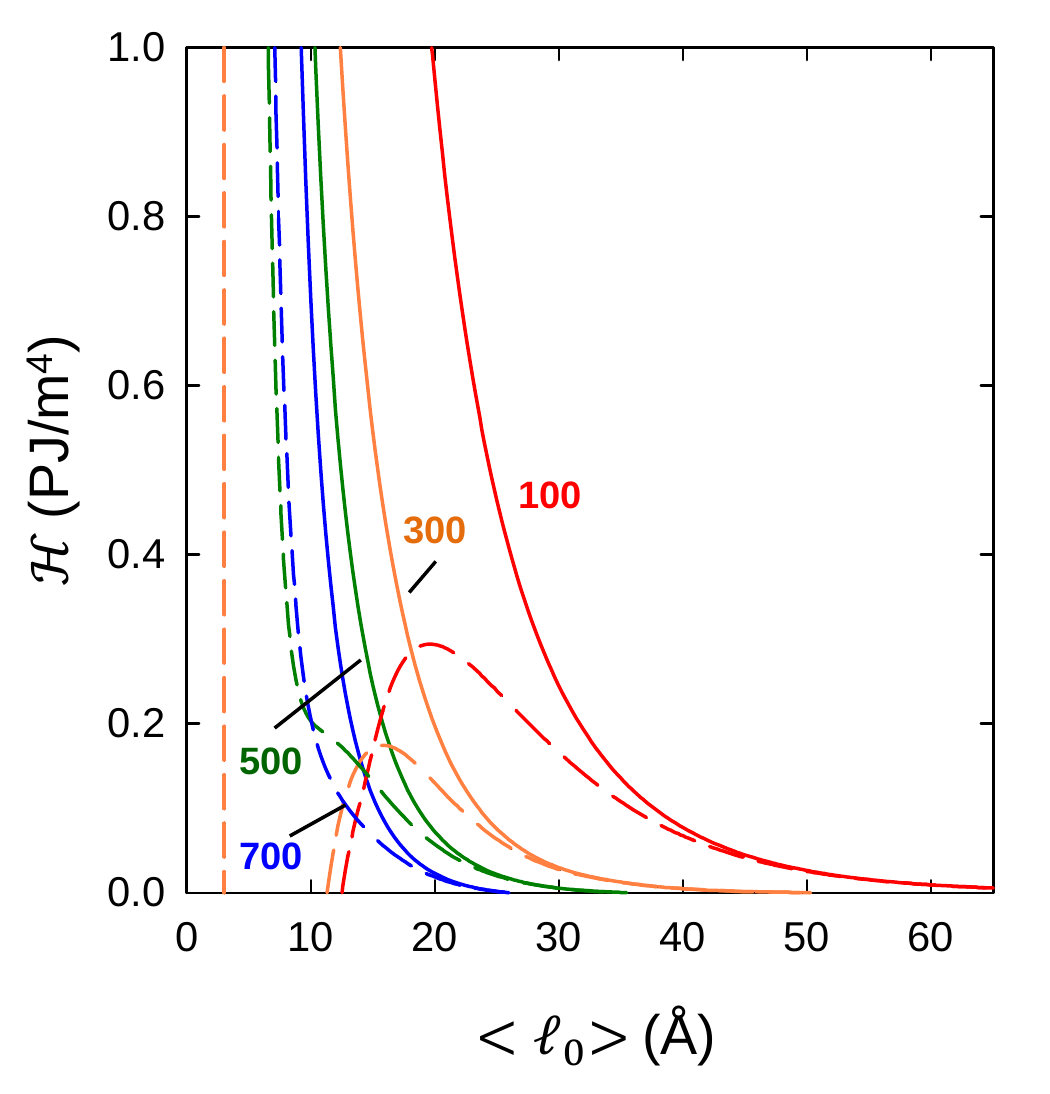}
	\caption{Behavior of the Hessian  $ \mathcal{H}\equiv \partial^2 f_{{\rm var}}/\partial \ell_0^2 $ as a function of membrane equilibrium spacing $\langle \ell_0 \rangle$ and salt concentrations for CC (solid lines) and CP (dashed lines) boundary conditions. Colors represent different salt concentrations, given in units of mM by numbers next to the data. All the CC curves show a non-negative Hessian and the Gaussian theory on which they are based, is stable. The CP curves show a more complicated behavior often exhibiting negative values of the Hessian, signaling a breakdown of the Gaussian theory.}
\label{fig:fluct_oscillations}
\end{figure*}

Solid and dashed lines in Figure~\ref{fig:flucts} show the results of the joint analysis of bilayer separations and fluctuations in terms of the model described in the previous section. As already indicated, this creates stringent constraints on the accuracy of the fits and imposes severe consistency checks on the theoretical models. Nevertheless, even within the confines of these constraints, we find a good agreement with experimental data. In addition, the analysis was performed concurrently with a validity check for the applied variational Gaussian approximation, a check altogether different and independent from the two mentioned above. As detailed in the previous section, the proper function for performing this test is provided by the Hessian,  Eq. \ref{gfdyei}, shown in Figure~\ref{fig:fluct_oscillations} for the two extreme cases of boundary conditions. The Hessian needs to be non-negative in order that the Gaussian theory be valid, which is clearly the case for the CC boundary conditions. On the other hand, the CP boundary condition does not in general conform to this. 

In summary, the fitting procedure to which we subject the data is unparalleled in its usual implementations, constrained first of all by the required consistency between the osmotic pressure and the fluctuations data, as well as the consistency of the theoretical model description.

For constant charge boundary conditions, $\mathcal{H}$ dropped monotonously with increasing bilayer separations for all salt concentrations and became negative for sufficiently large $ \langle \ell_0 \rangle$ (e.g. $ \langle \ell_0 \rangle > 27$ \AA\/ (700 mM), $ \langle \ell_0 \rangle > 36$ \AA\/ (500 mM) salt, $ \langle \ell_0 \rangle > 50$ \AA\/ for (300 mM)). This means that the variational Gaussian approximation is not reliable at such separations, which occurs, however, just outside the experimentally probed range (Fig.~\ref{fig:flucts}). $\mathcal{H}$ for constant potential boundary conditions approached that of constant charge boundary conditions for large separations. However, the Hessian of $c_{\rm NaCl} = 100$ and 300 mM exhibited a range of negative values at smaller $ \langle \ell_0 \rangle$-values. As our model assumes that membranes of the same type (e.g., DPPG) exhibit the same BC, we consequently have to rule out the constant potential BC in the interpretation of our data.

Henceforth, we discuss results for the interaction parameters for constant charge BC only. 
Within experimental uncertainty we found the surface charge density to be  constant $\sigma_s \sim 2.40 \times 10^{-3}$ e/\AA$^2$ for $c_{\rm NaCl} \le 500$ mM, decreasing slightly at 700 mM salt to $2.29 \times 10^{-3}$ e/\AA$^2$. The penetration coefficient also emerged as constant, $c \sim 0.1$, and not to change with salt concentration. The latter result is consistent with our SDP SAXS data analysis, which showed no change in DPPG membrane structure in the presence of NaCl. Likewise, the hydration pressure amplitudes $P_H$ ($\sim 1974$~atm) and the corresponding decay constant $\lambda_H$ ($\sim 1$~\AA)  exhibited values which are well within the reported ranges~\cite{Parsegian.1995,Petrache.1998b,Kollmitzer.2015b}, and also did not change significantly with the salt content. The amplitude $P_{es}$ shows a slight decrease upon increasing salt concentration as reported in Tab.~\ref{table:tablefluct1}, which is however well below the overall accuracy of the fits and the data.

\begin{table} [h]
\begin{center}
    \begin{tabular}{|c|c|c|c|}
    \hline
    $c_{\rm NaCl}$ (mM) &
    $\lambda_D$ (\AA) &
    $P_{es}$ ($10^{-4}$ $k_{{\rm B}}T/$ \AA$^{3}$) & 
    $W$ ($k_{\rm B} T$)
     
    \\ \hline
    
    $100$ & $9.46$ & $5.3$ & $1.8$   \\ \hline
     
    $300$ & $5.46$ & $5.3$ & $1.8$  \\ \hline
    
    $500$  & $4.23$ & $5.1$ & $1.3$  \\ \hline
    
    $700$ & $3.58$ & $4.9$ & $1.3$   \\ \hline
        \end{tabular}
        \caption{Interaction parameters Debye screening length, $\lambda_D$, electrostatic pressure amplitude, $P_{es}$ and the Hamaker coefficient, $W$, in DPPG bilayers as a function of NaCl concentration in the bathing solution. Errors for $P_{es}$ and $W$ are estimated to be 15-20\%, the largest fraction of this error due to the vestigial osmotic pressure and the consequent absence of solid zero pressure data.} 
        \label{table:tablefluct1}
\end{center}
\end{table}

\begin{figure}
		\includegraphics[width=0.7\textwidth]{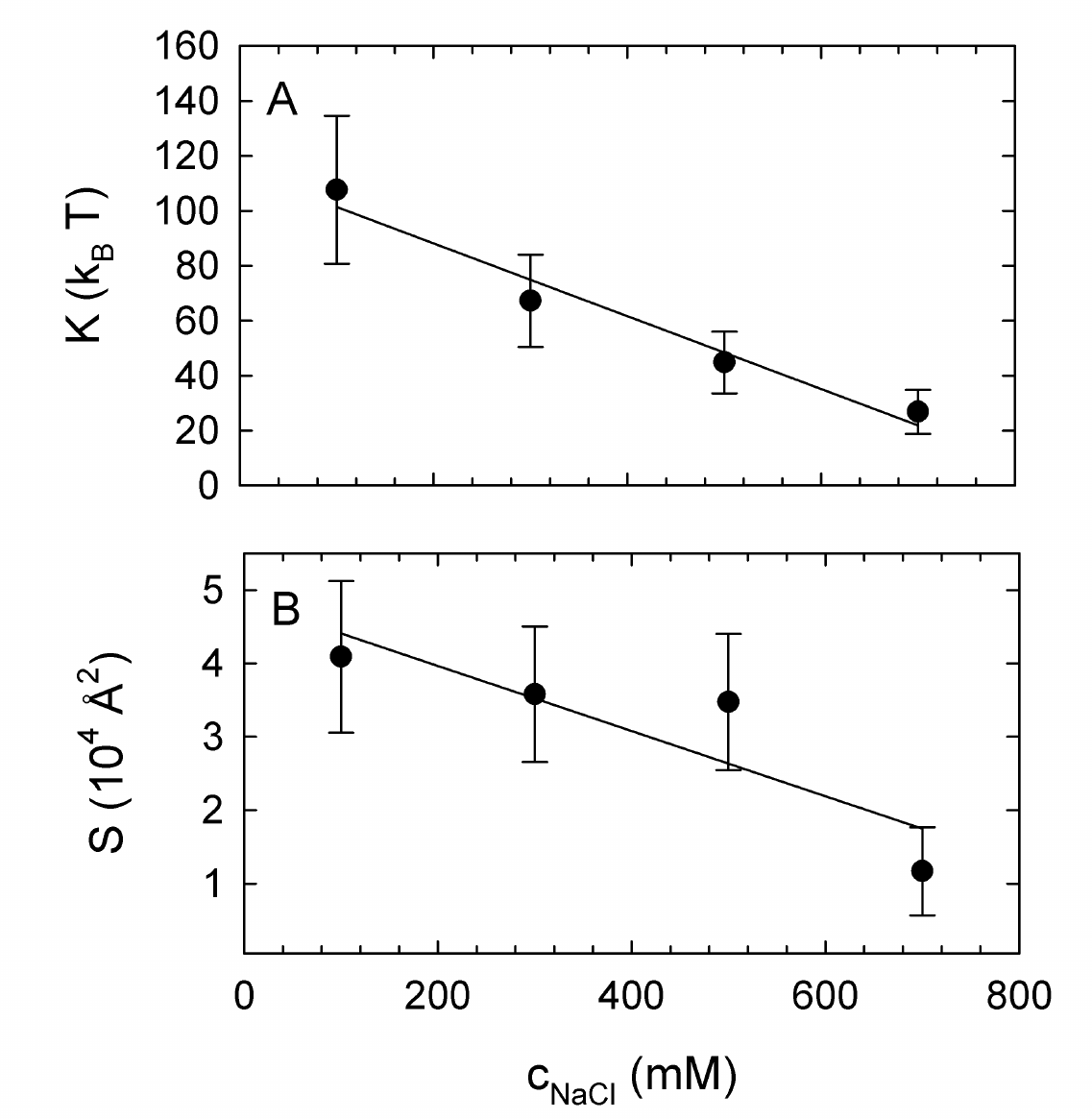}
	\caption{Bending rigidity (panel A) and projected membrane surface area (panel B) as a function of salt concentration. The bending rigidity clearly attests to the softening of the bilayer elasticity due to solution ion screening, leading consistently also to a diminished projected membrane surface area.}
\label{fig:params}
\end{figure}

Unlike previous reports~\cite{Petrache.2006b,Petrache.2006} we were, however, not able to find a significant change of the Hamaker coefficient with salt due to very large experimental uncertainties in determining this fitting parameter. They are a consequence of the fact that the value of the Hamaker coefficient in the fitting procedure relies on a very few data points at very small pressures,  as is clearly evident from Fig.~\ref{fig:flucts}. The same feature would be exhibited also in already published data sets, where the equilibrium spacing is inferred from typically just a few data points at small osmotic pressures. These uncertainties are, however, seldom clearly pointed out and discussed. Finally, the bending rigidity, which was assumed to remain constant in the previous work ~\cite{Petrache.2006b,Petrache.2006} did change significantly upon increasing NaCl concentration (Fig.~\ref{fig:params}A) with a slope of $-0.13 \pm 0.02$ $k_{\rm B} T/\rm mM$. The high fidelity of these results are a distinctive feature of our analysis that takes properly into account also bilayer shape fluctuations, which are coupled to the bilayer elasticity. Finally, to the best of our knowledge, this is the first experimental evidence for softening of charged lipid bilayers by ion screening. 

At 700 mM NaCl concentration, $K$ is on the same order of magnitude as bending rigidities reported for uncharged membranes (see, e.g., ref. ~\cite{Rappolt.2008,Dimova.2014}). At 100 mM salt in turn, the bending rigidity of DPPG is similar to a gel phase of charge neutral bilayers, which were reported to be about five times more rigid than fluid bilayers~\cite{Lee.2001}. Large $K$-values are expected for charged bilayers by theory ~\cite{Andelman.2006} and have been reported previously by Mertins \textit{et al.}~\cite{Mertins.2013} for dioleoyl phosphatidylcholine/dioleoyl phosphatidylglycerol mixtures (88 $k_{{\rm B}} T$) or chitosan coated dioleoyl phosphatidylcholine vesicles (72 $k_{{\rm B}} T$), but there typically is a large spread of bending rigidities from different experimental techniques.~\cite{Rappolt.2008}

Finally, we note that the projected membrane area $S$ decreases from $\sim 4.1 \times 10^4 {\rm \AA}^2$ to $\sim 1.2 \times 10^4 {\rm \AA}^2$ in the studied NaCl concentration range with a slope of $-44 \pm 17$ \AA$^2$/mM (Fig.~\ref{fig:params}B). This is in line with our physical expectation that with increasing salt concentration, softening the bilayer elasticity, the undulations become stronger, resulting in the membrane becoming more crumpled and thus having a smaller projected area.  


\section{Conclusions}

We analyzed controlled swelling experiments performed on anionic lipid multilayers in a monovalent salt bathing solution. For osmotic pressures, defined by the added PEG osmolyte, we measured the small angle X-ray scattering accessible Bragg spacings as well as the fluctuation amplitudes of the first Bragg peak for constitutive bilayers in the multilamellar sample. The swelling behavior of charged anionic bilayers in salt solutions for four different concentrations of monovalent salt (NaCl), was fitted to the osmotic pressure dependencies of the mean bilayer separation and fluctuation amplitude predictions based on a theoretical model that hinges on a variant of the Feynman-Kleinert variational theory. From these measurements and fits, we found that the equilibrium interlamellar separation decreases with increasing salt concentration, which is expected on grounds that the electrostatic repulsion favoring multilayer swelling is reduced by the Debye ES screening. Furthermore, the fluctuations are indeed stronger for (i)~larger salt concentration (for the same interlamellar separation) and/or (ii)~larger interlamellar separation (for the same salt concentration), again owing to a reduced electrostatic repulsion.

One of the chief novelties and advantages of our theoretical approach is that it enables us to make concurrent and self-consistent predictions for both the equilibrium interlamellar separation and the fluctuation amplitude, experimentally accessible from osmotic stress experiments with osmotic pressure defined by the osmolyte (PEG) in the bulk solution, in combination with high-resolution x-ray scattering, enabling a joint refinement of the model parameters against the data. This is a definite advance over previous theoretical fitting models that only allow for fitting of the osmotic pressure but not the fluctuation amplitude. Furthermore, by making use of a variational Gaussian approximation~\cite{Lu.2015}, our theoretical approach accounts for the renormalization of the interaction potential by the presence of fluctuations, and thus does \emph{not} assume that the soft interaction potentials, either of the  fluctuation Helfrich type or of the soft DLVO type, are additive. This is again in contrast to previous fitting approaches that make {\sl a priori} assumptions about the additivity of the Helfrich fluctuation and interaction potentials, and do not account for the fluctuation renormalization of the interaction potentials~\cite{Petrache.1998b,Petrache.2006b}. A third novelty is our use of the Hessian within the framework of the variational Gaussian approximation as a tool for identifying an appropriate electrostatic boundary condition. On this basis, we were able to compare results from a model with constant charge BC and another with constant potential BC, and conclude that the constant charge BC gives a better approximation. 

From the analysis presented we were able to gain fundamental insights into the  interactions between anionic DPPG bilayers and their modulation by monovalent salt solutions. In particular, we highlight the decrease of bending rigidity by a factor of $\sim 4$ upon increasing salt concentration, which is expected on theoretical grounds~\cite{Andelman.2006} but has to be best of our knowledge not been reported from experiment before. Considering the large gradients of ions across cellular membranes this gives salts a significant role in modulating not only electrostatic interactions between adjacent membranes or membrane and any other macro molecule, but also membrane elasticity and thereby steric (fluctuation driven) interactions. This effect is possibly even more expressed for polyvalent ions and is currently explored in our laboratories.

Last but not least, we identified a {\sl vestigial osmotic pressure} in the  completely unstressed state with a nominally zero osmotic pressure, that resulted in a displaced, i.e., diminished, equilibrium spacing. We proposed that it originates in the pealing off of small unilamellar vesicles into the bathing solution that then  act as an additional source of osmotic stress not set by the osmolyte (PEG). The details of this process would be difficult to quantify and are possibly of non-equilibrium origin.

\section{Acknowledgement}

The authors thank Heinz Amenitsch for valuable experimental assistance at the Austrian SAXS beamline. This work was financially supported by the Agency for research and development of Slovenia (ARRS) and the Austrian Science Funds (FWF) under the bilateral SLO-A Grant Nos. N1-0019 (RP) and I1304-B20 (GP), respectively.

\section{Supporting Information}

In the Supporting Information, we fill in the requisite calculational steps and show the more lengthy formulas  we used to perform the modeling and analysis of experimental data.   
We begin by explaining how  to transform the problem of a two-membrane system into that of a membrane fluctuating next to a hard wall (cf. Sec. II B). Next, we derive the variational equation used to approximate the mean square undulation amplitude, and obtain a corresponding expression for the osmotic pressure. 
Finally, we show how the quantity $\Delta_{{\rm full}}$ corresponds to the fluctuation amplitude $\Delta_{{\rm exp}}$ measured in experiment. This quantity involves both a zero mode fluctuation contribution and an undulatory contribution. We give explicit formulas for the computation of both contributions. 

\begin{figure}
		\includegraphics[width=0.7\textwidth]{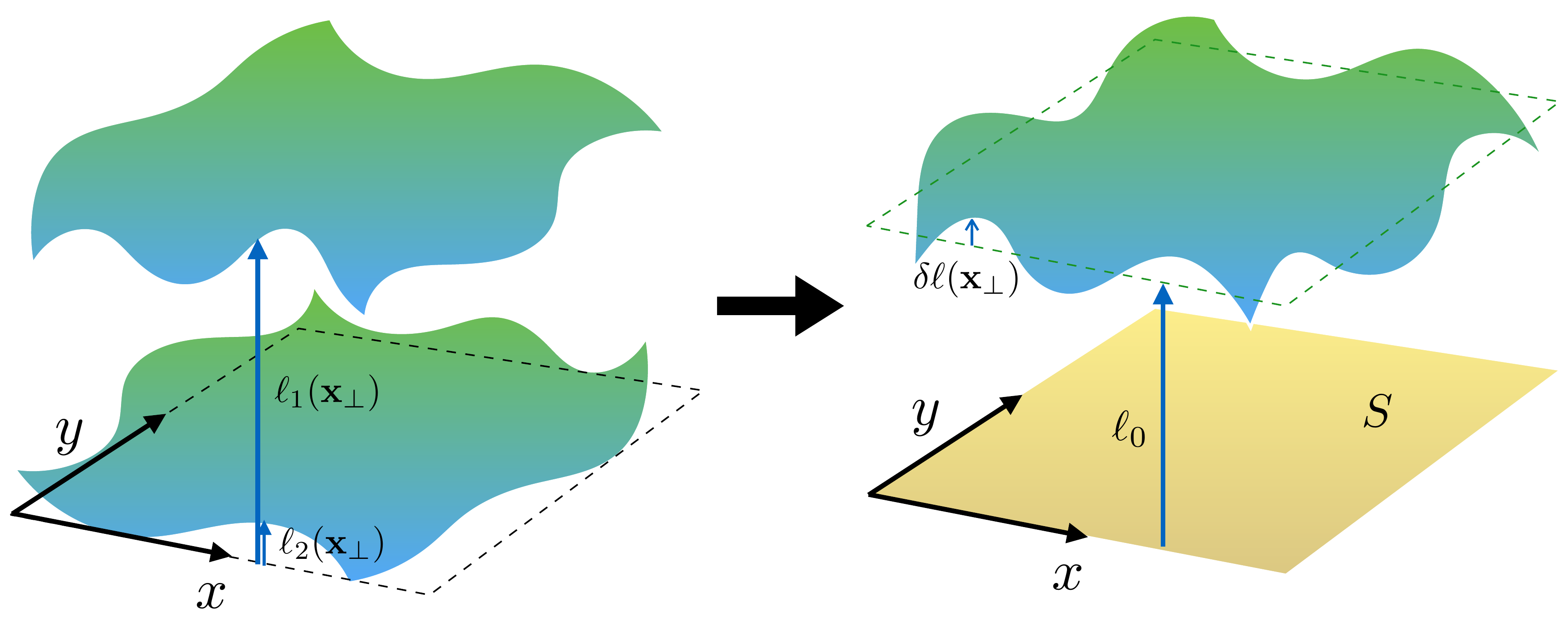}
	\caption{Schematic of the $\ell_1, \ell_2 \longrightarrow \ell$ transformation used in data modeling. On the left, the two membranes are on average co-planar with one another, and each membrane is also co-planar with a base reference plane (shown with the black dashed line) which has two-dimensional coordinates $\mathbf{x}_\perp = (x,y)$. The local height of the upper (lower) membrane relative to the base plane is $\ell_1(\mathbf{x}_\perp)$ ($\ell_2(\mathbf{x}_\perp)$). On the right, we make the passage to a relative height $\ell(\mathbf{x}_\perp) = \ell_1(\mathbf{x}_\perp)-\ell_2(\mathbf{x}_\perp)$. In this new representation, the lower membrane becomes a flat wall (shown in beige color). The upper membrane, which is not the same as the upper membrane before the transformation, is on average located at a height $\ell_0$ above the wall, with a local undulation $\delta\ell(\mathbf{x}_\perp)$ about the average height, such that $\ell(\mathbf{x}_\perp) = \ell_0 + \delta\ell(\mathbf{x}_\perp)$. The cross-sectional area of the membrane projected onto the base plane is $S$.}\label{fig:lipowsky_trans}
\end{figure}

The bending energy of the two-membrane system is given by 
\begin{equation}
H_{b} = \frac{1}{2} \int \! d^2\mathbf{x}_\perp \big[ K_1 (\nabla_\perp^2\ell_1(\mathbf{x}_\perp))^2 + K_2 (\nabla_\perp^2\ell_2(\mathbf{x}_\perp))^2 \big],
\end{equation}
where $\mathbf{x}_\perp=(x,y)$ is a two-dimensional coordinate on the transverse projected plane of the membranes, $\nabla_\perp \equiv (\partial_x,\partial_y)$ is a two-dimensional gradient operator acting in the transverse plane, $\ell_1(\mathbf{x}_\perp)$ and $\ell_2(\mathbf{x}_\perp)$ are the local ``heights" of the first and second membranes measured with respect to a reference plane, and $K_1$ and $K_2$ are the bending rigidities of the membranes. Instead of using $\ell_1$ and $\ell_2$ we find it more convenient to work with the ``center-of-mass" and relative coordinates $\ell_c$ and $\ell$; we define these by $\ell_c \equiv (K_1 \ell_1 + K_2 \ell_2)/(K_1 + K_2)$,  $\ell \equiv \ell_2 - \ell_1$ and $K_{{\rm eff}} \equiv (K_1 K_2)/(K_1 + K_2)$, where $K_{{\rm eff}}$ is the ``effective" bending rigidity. For similar membranes, $K_1 = K_2 \equiv K$, and the bending rigidity of a single membrane is thus related to the effective bending rigidity via $K = 2 K_{{\rm eff}}$. In terms of our new coordinates, the bending energy becomes
\begin{equation}
H_{b} = \frac{K_{{\rm eff}}}{2} \int \! d^2\mathbf{x}_\perp \, (\nabla^2 \ell(\mathbf{x}_\perp))^2,
\end{equation}
where we have neglected nonlinear gradient coupling terms  that are less important than the coupling terms stemming from the interaction potential.~\cite{Lipowsky.1986}  We can thus interpret this bending energy as describing the steric interaction between a fluctuating membrane and a hard planar wall (see Fig.~\ref{fig:lipowsky_trans}).  

Next, we derive the variational equation used in Sec. II B. As explained in Ref.~\cite{Lu.2015}, we can obtain a variational approximation to the mean square undulation $\Delta_u^2$ of the membrane for a given value of $\ell_0$ by varying the variational free energy $f_{var}$ with respect to $\Delta_u$ for that value of $\ell_0$~\footnote{In Ref.~\cite{Lu.2015} the mean square undulation was denoted by $\sigma^2$.}: 
\begin{equation}
\label{eq:varequation}
\frac{\partial w_u}{\partial\Delta_u} - \frac{(k_{{\rm B}}T)^2}{64 K_{{\rm eff}} \Delta_u^3} + \frac{9 k_{{\rm B}}T c^2 \Delta_u}{4\ell_0^4} =0.
\end{equation}
Here $w_u$ is the interaction potential per unit area~\footnote{$w_u$ is called $\widetilde{w}_{\sigma^2}$ in Ref.~\cite{Lu.2015}.}. 
For a system subject to \emph{constant charge} boundary condition (BC) $w_u$ is given by 
\begin{eqnarray}
w_u
&=& P_H \lambda_H e^{-\frac{\ell_0}{\lambda_H}+\frac{\Delta_u^2}{2\lambda_H^2}}
- W\Delta_u^2 g(\ell_0) M_4(\ell_0) 
- \frac{1}{3} W g(\ell_0) M_2(\ell_0) 
\nonumber\\
&&+
P_{es} \lambda_D (\coth(\ell_0/2\lambda_D)-1)  
+
\frac{P_{es} \Delta_u^2 \sinh(\ell_0/\lambda_D)}{8\lambda_D (\sinh(\ell_0/2\lambda_D))^4},
\end{eqnarray}
where $g(\ell_0)$ is a cut-off function~\cite{Podgornik.2004}, defined by  
\begin{equation}
g(\ell_0) \equiv \Big(1 - e^{-\frac{\ell_0^2}{2\lambda_H^2}}\Big)^3,
\end{equation}
and $M_n(\ell_0) \equiv \big[ \frac{1}{\ell_0^n} -\frac{2}{(\ell_0+d_B)^n} +\frac{1}{(\ell_0+2d_B)^n} \big]/4\pi$.  

For \emph{constant potential} BC $w_u$ is given by 
\begin{eqnarray}
w_u &=& 
P_H \lambda_H e^{-\frac{\ell_0}{\lambda_H}+\frac{\Delta_u^2}{2\lambda_H^2}}
- W\Delta_u^2 g(\ell_0) M_4(\ell_0) 
- \frac{1}{3} W g(\ell_0) M_2(\ell_0)
\nonumber\\
&&-
P_{es} \lambda_D (\tanh(\ell_0/2\lambda_D)-1)  
+
\frac{2 P_{es} \Delta_u^2 (\sinh(\ell_0/2\lambda_D))^4}{\lambda_D (\sinh(\ell_0/\lambda_D))^3}.
\end{eqnarray}
Substituting the above expressions for $\widetilde{w}$ into Eq.~(\ref{eq:varequation}) yields
\begin{equation}
\label{eq:sigma-variation}
-\frac{(k_{{\rm B}}T)^2}{128 K_{{\rm eff}} \Delta_u^4} + \frac{9 k_{{\rm B}}T c^2}{8\ell_0^4} 
+ \frac{P_H}{2 \lambda_H} e^{-\frac{\ell_0}{\lambda_H}+\frac{\Delta_u^2}{2\lambda_H^2}}
+ \frac{P_{es}}{2 \lambda_D} f_e(\ell_0)
- W g(\ell_0) M_4(\ell_0)
=0, 
\end{equation}
where  $f_e(\ell_0)$ is a BC-dependent function defined by
\begin{eqnarray}
f_e(\ell_0) 
\equiv 
\left\{ \begin{array}{ll} 
\frac{\sinh(\ell_0/\lambda_D)}{4 (\sinh(\ell_0/2\lambda_D))^4} & 
\quad \mbox{(constant charge BC)}      
   \vspace{3mm}\\
\frac{4 (\sinh(\ell_0/2\lambda_D))^4}{(\sinh(\ell_0/\lambda_D))^3} & 
\quad \mbox{(constant potential BC)}.   
\end{array} \right.   
\end{eqnarray}
The solution to $\Delta_u^2$ in Eq.~(\ref{eq:sigma-variation}) then gives us the variational approximation to the mean square undulation. 



An expression for the effective osmotic pressure is obtained by varying $f_{{\rm var}}$ with respect to $\ell_0$: 
\begin{eqnarray}
P_{osm} &=& 
\frac{9 k_{{\rm B}}T c^2 (\Delta_u(\ell_0^\ast))^2}{2(\ell_0^\ast)^5} + \frac{3c^2 k_{{\rm B}}T}{4 (\ell_0^\ast)^3}
+ P_H e^{-\frac{\ell_0^\ast}{\lambda_H}+\frac{(\Delta_u(\ell_0^\ast))^2}{2\lambda_H^2}} 
+ P_{es} h_e(\ell_0^\ast)
\\
&&
- 4 W (\Delta_u(\ell_0^\ast))^2 g(\ell_0^\ast) M_5(\ell_0^\ast)
-
\frac{2}{3} W g(\ell_0^\ast) M_3(\ell_0^\ast)
+
\frac{1}{3} W g'(\ell_0^\ast) \big[ M_2(\ell_0^\ast) + 3 (\Delta_u(\ell_0^\ast))^2 M_4(\ell_0^\ast) \big],
\nonumber
\end{eqnarray}
where $\ell_0^\ast$ is a saddle-point approximation to the equilibrium rigid bilayer separation $\langle \ell_0 \rangle$, and 
\begin{eqnarray}
h_e(x) 
\equiv 
\left\{ \begin{array}{ll} 
\frac{((\Delta_u(x))^2+2\lambda_D^2)\cosh(x/\lambda_D) + 2((\Delta_u(x))^2-\lambda_D^2)}{8 \lambda_D^2 (\sinh(x/2\lambda_D))^4}
 & 
\quad \mbox{(constant charge BC)}      
   \vspace{3mm}\\
\frac{((\Delta_u(x))^2+2\lambda_D^2)\cosh(x/\lambda_D) - 2((\Delta_u(x))^2-\lambda_D^2)}{8 \lambda_D^2 (\cosh(x/2\lambda_D))^4} & 
\quad \mbox{(constant potential BC)}.   
\end{array} \right.   
\end{eqnarray}
Next, we consider the fluctuation amplitude determined by experiment. It is expressed by the formula~\cite{Petrache.1998b}:
\begin{equation}
\Delta_{{\rm exp}} = \sqrt{\big\langle \big[ u(\mathbf{x}_\perp, n+1) - u(\mathbf{x}_\perp, n) \big]^2 \big\rangle}.
\end{equation}
As explained in the paper, this formula originates from a smectic model of lipid bilayers, with $u(\mathbf{x}, n)$ representing thermal fluctuations from the equilibrium position of the $n$-th membrane in the direction perpendicular to the average plane of the membrane.
The corresponding quantity in our two-membrane model is given by 
\begin{eqnarray}
\Delta_{{\rm full}} &=& \sqrt{\big\langle \big[ \ell_2 - \langle \ell_2 \rangle - (\ell_1 - \langle \ell_1 \rangle) \big]^2 \big\rangle} 
\nonumber\\
&=& \sqrt{\langle ( \ell - \langle \ell \rangle)^2 \rangle} 
\end{eqnarray}
Further writing $\ell = \langle \ell_0 \rangle + \delta\ell_0 + \delta\ell$, we arrive at 
\begin{equation}
\label{eq:fullfluct}
\Delta_{{\rm full}} = \sqrt{\langle \delta\ell_0^2 \rangle + \langle \delta\ell^2 \rangle}
\end{equation}
Implicit in the thermal averaging $\langle \dots \rangle$ is an averaging over the membrane's projected area, so that $\langle \delta \ell \rangle = 0$ and $\langle \delta \ell_0 \rangle = 0$. Finally, we approximate $\langle \delta\ell^2 \rangle$ by $\Delta_u^2$ and $\langle \delta\ell_0^2 \rangle$ by Eq.~(\ref{eq:meansquarefluct}).

The calculation of $\Delta_{{\rm full}}$ requires us to calculate $\langle \delta\ell_0^2 \rangle$. To this end we consider the partition function in the variational approximation:
\begin{equation}
Z \approx \int \! d\ell_0 \, e^{-\beta S f_{var}(\ell_0)}
\end{equation}
Note that the $\sigma$'s that occur in $f_{var}$ are functions of $\ell_0$, as we have already approximated $Z$ by the optimized contribution of $\sigma$ (or $\delta\ell$) for each value of $\ell_0$; mathematically, such an optimization is expressed by Eq.~(\ref{eq:varequation}). 
To find $\langle \delta\ell_0^2 \rangle$, we expand the exponent to second order in $\delta\ell_0$ around the saddle-point value $\ell_0^\ast$: 
\begin{equation}
Z \approx \int \! d\ell_0 \, e^{-\beta S f_{var}(\ell_0^\ast) -\frac{1}{2}\beta S \big[ \frac{\partial^2 f_{var}}{\partial \ell_0^2} \big]_{\ell_0^\ast} \delta\ell_0^2}
\end{equation}
Thus we see that $\langle \delta\ell_0^2 \rangle$ can be approximated by
\begin{equation}
\label{eq:meansquarefluct}
\langle \delta\ell_0^2 \rangle = 
\frac{k_{\rm{B}} T}{S} 
\left[ \frac{\partial^2 f_{var}}{\partial \ell_0^2} \right]_{\ell_0^\ast}^{-1}
\end{equation}
The second derivative $\frac{\partial^2 f_{var}}{\partial \ell_0^2}$ is given by
\begin{equation}
\frac{\partial^2 f_{var}}{\partial \ell_0^2} = \frac{\partial^2 w_u}{\partial\ell_0^2} + \frac{45 k_{\rm{B}}T c^2 \Delta_u^2}{2\ell_0^6} + \frac{9 k_{\rm{B}} T c^2}{4\ell_0^4} + \left( \frac{\partial^2 w_u}{\partial\Delta_u \partial\ell_0} - \frac{9 k_{\rm{B}} T c^2 \Delta_u}{\ell_0^5} \right) \frac{\partial\Delta_u}{\partial\ell_0}
\end{equation}
In deriving this expression we have made use of the fact that $\sigma=\sigma(\ell_0)$ through Eq.~(\ref{eq:varequation}). 
The quantity $\partial \Delta_u / \partial \ell_0$ can be computed by differentiating Eq.~(\ref{eq:varequation}) with respect to $\ell_0$, which yields 
\begin{equation}
\frac{\partial \Delta_u}{\partial \ell_0} = \frac{\frac{9 k_{\rm{B}} T c^2 \Delta_u}{\ell_0^5} - \frac{\partial^2 w_u}{\partial\Delta_u \partial\ell_0}}{\frac{\partial^2 w_u}{\partial\Delta_u^2} + \frac{3(k_{\rm{B}} T)^2}{64 K_c \Delta_u^4} + \frac{9 k_{\rm{B}} T c^2}{4 \ell_0^4}}
\end{equation}
We thus have 
\begin{equation}
\label{eq:fullgory}
\langle \delta\ell_0^2 \rangle \approx 
\frac{k_{\rm{B}} T}{S} 
\left[
\frac{\partial^2 w_u}{\partial\ell_0^2} + \frac{45 k_{\rm{B}}T c^2 \Delta_u^2}{2\ell_0^6} + \frac{9 k_{\rm{B}} T c^2}{4\ell_0^4} -
\frac{\left( \frac{9 k_{\rm{B}} T c^2 \Delta_u}{\ell_0^5} - \frac{\partial^2 w_u}{\partial\Delta_u \partial\ell_0} \right)^2}{\frac{\partial^2 w_u}{\partial\Delta_u^2} + \frac{3(k_{\rm{B}} T)^2}{64 K_c \Delta_u^4} + \frac{9 k_{\rm{B}} T c^2}{4 \ell_0^4}}
\right]_{\ell_0^\ast}^{-1}
\end{equation}
For constant charge BC, the derivatives of $w_u$ are given by 
\begin{subequations}
\begin{eqnarray}
\frac{\partial^2 w_u}{\partial \ell_0^2} &=& 
-\frac{W}{2\pi} \Big(1 - e^{-\frac{\ell_0^2}{2\lambda_H^2}}\Big) 
\bigg\{  
\Big(1 - e^{-\frac{\ell_0^2}{2\lambda_H^2}}\Big)^2 
\left[ \frac{1}{\ell_0^4} - \frac{2}{(\ell_0+d_B)^4} + \frac{1}{(\ell_0+2d_B)^4} \right]
\nonumber\\
&&\quad- 
\frac{2 \ell_0}{\lambda_H^2} \Big(1 - e^{-\frac{\ell_0^2}{2\lambda_H^2}}\Big) e^{-\frac{\ell_0^2}{2\lambda_H^2}} 
\left[ \frac{1}{\ell_0^3} - \frac{2}{(\ell_0+d_B)^3} + \frac{1}{(\ell_0+2d_B)^3} \right]
\nonumber\\
&&\quad-
\frac{d_B^2(2d_B^2 + 6d_B \ell_0 + 3\ell_0^2)}{\ell_0^2(d_B+\ell_0)^2(2d_B + \ell_0)^2 \lambda_H^4} 
\Big[ \Big( 1 - 3 e^{-\frac{\ell_0^2}{2\lambda_H^2}} \Big) \ell_0^2 - \Big( 1 - e^{-\frac{\ell_0^2}{2\lambda_H^2}}  \Big) \lambda_H^2 \Big] 
e^{-\frac{\ell_0^2}{2\lambda_H^2}}
\bigg\}
\nonumber\\
&&+
\frac{W\Delta_u^2}{4 \pi} 
\Big(1 - e^{-\frac{\ell_0^2}{2\lambda_H^2}}\Big) 
\bigg\{ - 20 \Big(1 - e^{-\frac{\ell_0^2}{2\lambda_H^2}}\Big)^2 
\left[ \frac{1}{\ell_0^6} - \frac{2}{(\ell_0+d_B)^6} + \frac{1}{(\ell_0+2d_B)^6} \right] 
\nonumber\\
&&\quad+
\frac{24\ell_0}{\lambda_H^2} 
\Big(1 - e^{-\frac{\ell_0^2}{2\lambda_H^2}}\Big) e^{-\frac{\ell_0^2}{2\lambda_H^2}}
\left[ \frac{1}{\ell_0^5} - \frac{2}{(\ell_0+d_B)^5} + \frac{1}{(\ell_0+2d_B)^5} \right] 
\nonumber\\
&&\quad+
\frac{3}{\lambda_H^4}
\Big(1 - e^{-\frac{\ell_0^2}{2\lambda_H^2}}\Big) e^{-\frac{\ell_0^2}{2\lambda_H^2}}
\left[ \frac{1}{\ell_0^4} - \frac{2}{(\ell_0+d_B)^4} + \frac{1}{(\ell_0+2d_B)^4} \right] 
\nonumber\\
&&\qquad\times
\Big[ \Big( 1 - 3 e^{-\frac{\ell_0^2}{2\lambda_H^2}} \Big) \ell_0^2 - \Big( 1 - e^{-\frac{\ell_0^2}{2\lambda_H^2}}  \Big) \lambda_H^2 \Big] 
\bigg\}
\nonumber\\
&&+ 
\frac{P_H e^{-\frac{\ell_0}{\lambda_H} + \frac{\Delta_u^2}{2\lambda_H^2}}}{\lambda_H}
+\frac{P_{es} \sinh(\frac{\ell_0}{\lambda_D})}{4 \lambda_D \sinh^4(\frac{\ell_0}{2\lambda_D})}
+
\frac{P_{es} \Delta_u^2 (11\cosh(\frac{\ell_0}{2\lambda_D})+\cosh(\frac{3\ell_0}{2\lambda_D}))}{16 \lambda_D^3 \sinh^5(\frac{\ell_0}{2\lambda_D})},
\end{eqnarray}
\begin{eqnarray}
\frac{\partial^2 w_u}{\partial \ell_0 \partial \Delta_u} 
&=& 
\frac{2W \Delta_u}{\pi} \Big( 1 - e^{-\frac{\ell_0^2}{2\lambda_H^2}} \Big)^3 
\left[ \frac{1}{\ell_0^5} - \frac{2}{(\ell_0+d_B)^5} + \frac{1}{(\ell_0+2d_B)^5} \right] 
\nonumber\\
&&-\frac{3 W \ell_0 \Delta_u}{2\pi \lambda_H^2}
\Big( 1 - e^{-\frac{\ell_0^2}{2\lambda_H^2}} \Big)^2 e^{-\frac{\ell_0^2}{2\lambda_H^2}} 
\left[ \frac{1}{\ell_0^4} - \frac{2}{(\ell_0+d_B)^4} + \frac{1}{(\ell_0+2d_B)^4} \right] 
\nonumber\\
&&-
\frac{\Delta_u P_H}{\lambda_H^2} e^{-\frac{\ell_0}{\lambda_H} + \frac{\Delta_u^2}{2\lambda_H^2}} 
+ 
\frac{P_{es} \Delta_u \cosh(\frac{\ell_0}{\lambda_D})}{4 \lambda_D^2 \sinh^4(\frac{\ell_0}{2\lambda_D})}
- 
\frac{P_{es} \Delta_u \sinh^2(\frac{\ell_0}{\lambda_D})}{4 \lambda_D^2 \sinh^6(\frac{\ell_0}{2\lambda_D})},
\\
\frac{\partial^2 w_u}{\partial \Delta_u^2} &=& 
-\frac{W}{2\pi} 
\left[ \frac{1}{\ell_0^4} - \frac{2}{(\ell_0+d_B)^4} + \frac{1}{(\ell_0+2d_B)^4} \right] 
+ 
\frac{P_H (\Delta_u^2 + \lambda_H^2)}{\lambda_H^3} e^{-\frac{\ell_0}{\lambda_H} + \frac{\Delta_u^2}{2\lambda_H^2}} 
\nonumber\\
&&+ 
\frac{P_{es} \sinh(\frac{\ell_0}{\lambda_D})}{4 \lambda_D \sinh^4(\frac{\ell_0}{2\lambda_D})}.
\end{eqnarray}
\end{subequations}

For constant potential BC, the derivatives of $w_u$ are given by 
\begin{subequations}
\begin{eqnarray}
\frac{\partial^2 w_u}{\partial \ell_0^2} 
&=& 
-\frac{W}{2\pi} \Big( 1 - e^{-\frac{\ell_0^2}{2\lambda_H^2}} \Big)^3 
\left[ \frac{1}{\ell_0^4} - \frac{2}{(\ell_0+d_B)^4} + \frac{1}{(\ell_0+2d_B)^4} \right] 
\nonumber\\
&&+
\frac{W \ell_0}{\pi \lambda_H^2}
\Big( 1 - e^{-\frac{\ell_0^2}{2\lambda_H^2}} \Big)^2 e^{-\frac{\ell_0^2}{2\lambda_H^2}} 
\left[ \frac{1}{\ell_0^3} - \frac{2}{(\ell_0+d_B)^3} + \frac{1}{(\ell_0+2d_B)^3} \right] 
\nonumber\\
&&+
\frac{W d_B^2 (2d_B^2+6d_B \ell_0 +3\ell_0^2)}{2\pi \lambda_H^4 \ell_0^2(d_B+\ell_0)^2(2d_B+\ell_0)^2} 
\Big( 1 - e^{-\frac{\ell_0^2}{2\lambda_H^2}} \Big)
e^{-\frac{\ell_0^2}{2\lambda_H^2}} 
\Big[ \Big( 1 - 3 e^{-\frac{\ell_0^2}{2\lambda_H^2}} \Big) \ell_0^2 - \Big( 1 - e^{-\frac{\ell_0^2}{2\lambda_H^2}} \Big) \lambda_H^2 \Big]
\nonumber\\
&&-
\frac{5W \Delta_u^2}{\pi} 
\Big( 1 - e^{-\frac{\ell_0^2}{2\lambda_H^2}} \Big)^3 
\left[ \frac{1}{\ell_0^6} - \frac{2}{(\ell_0+d_B)^6} + \frac{1}{(\ell_0+2d_B)^6} \right] 
\nonumber\\
&&+
\frac{6W \ell_0 \Delta_u^2}{\pi \lambda_H^2}
\Big( 1 - e^{-\frac{\ell_0^2}{2\lambda_H^2}} \Big)^2 
e^{-\frac{\ell_0^2}{2\lambda_H^2}}
\left[ \frac{1}{\ell_0^5} - \frac{2}{(\ell_0+d_B)^5} + \frac{1}{(\ell_0+2d_B)^5} \right] 
\nonumber\\
&&+
\frac{3W \Delta_u^2}{4 \pi \lambda_H^4}
\Big( 1 - e^{-\frac{\ell_0^2}{2\lambda_H^2}} \Big) 
e^{-\frac{\ell_0^2}{2\lambda_H^2}} 
\left[ \frac{1}{\ell_0^4} - \frac{2}{(\ell_0+d_B)^4} + \frac{1}{(\ell_0+2d_B)^4} \right] 
\nonumber\\
&&\quad\times
\Big[ \Big( 1 - 3 e^{-\frac{\ell_0^2}{2\lambda_H^2}} \Big) \ell_0^2 - \Big( 1 - e^{-\frac{\ell_0^2}{2\lambda_H^2}} \Big) \lambda_H^2 \Big]
\nonumber\\
&&+ 
\frac{P_H e^{-\frac{\ell_0}{\lambda_H} 
+ \frac{\Delta_u^2}{2\lambda_H^2}}}{\lambda_H^2}
+ \frac{4 P_{es} \sinh^4(\frac{\ell_0}{2\lambda_D})}{\lambda_D \sinh^3(\frac{\ell_0}{\lambda_D})}
+
\frac{P_{es} (\cosh(\frac{\ell_0}{\lambda_D})-5)\tanh(\frac{\ell_0}{2\lambda_D}) \Delta_u^2}{8 \lambda_D^3 \cosh^4(\frac{\ell_0}{2\lambda_D})}
\end{eqnarray}

\begin{eqnarray}
\frac{\partial^2 w_u}{\partial \ell_0 \partial \Delta_u} 
&=& 
\frac{2 W \Delta_u}{\pi} 
\Big( 1 - e^{-\frac{\ell_0^2}{2 \lambda_H^2}} \Big)^3
\left[ \frac{1}{\ell_0^5} - \frac{2}{(\ell_0+d_B)^5} + \frac{1}{(\ell_0+2d_B)^5} \right] 
\nonumber\\
&&-
\frac{3 W \ell_0 \Delta_u}{2 \pi \lambda_H^2} 
e^{-\frac{\ell_0^2}{2 \lambda_H^2}}
\Big( 1 - e^{-\frac{\ell_0^2}{2\lambda_H^2}} \Big)^2
\left[ \frac{1}{\ell_0^4} - \frac{2}{(\ell_0+d_B)^4} + \frac{1}{(\ell_0+2d_B)^4} \right] 
\nonumber\\
&&-
\frac{P_H \Delta_u}{\lambda_H^2} 
e^{-\frac{\ell_0}{\lambda_H} + \frac{\Delta_u^2}{2\lambda_H^2}} 
+ 
\frac{P_{es} \Delta_u}{\lambda_D^2 \cosh^2(\frac{\ell_0}{2 \lambda_D})}
-
\frac{3 P_{es} \Delta_u \cosh(\frac{\ell_0}{\lambda_D})}{4 \lambda_D^2 \cosh^4(\frac{\ell_0}{2 \lambda_D})},
\\
\frac{\partial^2 w_u}{\partial \Delta_u^2} 
&=& 
-\frac{W}{2\pi} 
\Big( 1 - e^{-\frac{\ell_0^2}{2\lambda_H^2}} \Big)^3
\left[ \frac{1}{\ell_0^4} - \frac{2}{(\ell_0+d_B)^4} + \frac{1}{(\ell_0+2d_B)^4} \right] 
+ 
\frac{P_H (\Delta_u^2 + \lambda_H^2)}{\lambda_H^3} e^{-\frac{\ell_0}{\lambda_H} 
+ \frac{\Delta_u^2}{2\lambda_H^2}} 
\nonumber\\
&&
+ \frac{4 P_{es} \sinh^4(\frac{\ell_0}{2 \lambda_D})}{\lambda_D \sinh^3(\frac{\ell_0}{\lambda_D})}.
\nonumber\\
\end{eqnarray}
\end{subequations}
We determine the behavior of the full fluctuation amplitude by using the above expressions for the derivatives in conjunction with Eqs.~(\ref{eq:fullfluct}) and (\ref{eq:fullgory}). 

\bibliography{monovalent_salt}
	\bibliographystyle{plain}

\end{document}